\documentclass[a4paper,fleqn]{cas-dc}

\usepackage[authoryear,longnamesfirst]{natbib}

\def\tsc#1{\csdef{#1}{\textsc{\lowercase{#1}}\xspace}}
\tsc{WGM}
\tsc{QE}

\begin{document}
\let\WriteBookmarks\relax
\def\floatpagepagefraction{1}
\def\textpagefraction{.001}

\shorttitle{Searching for new heavy fermions with deep learning}    

\shortauthors{S.V. Dordevic}  

\title [mode = title]{Searching for new heavy fermions with deep learning}  



%

\author[1]{S.V. Dordevic}[orcid=0000-0001-8360-7577]



\ead{dsasa@uakron.edu}



\affiliation[1]{organization={Department of Physics, The University of Akron},
            city={Akron},
            postcode={44325}, 
            state={OH},
            country={USA}}







\cortext[1]{S.V. Dordevic}



\begin{abstract}
Deep learning models were developed and implemented 
to aid the search for new heavy fermion compounds. For the purpose
of these calculations a database of more than 200 heavy fermions
was compiled from the literature. The deep learning networks trained on the 
database were then used
for regression calculations, and predictions were made about the 
coherence temperature, Sommerfeld coefficient and carrier effective mass
of potential new heavy fermions. 
Classification calculations were also performed in order to check whether
predicted heavy fermions are superconducting and/or antiferromagnetic.  
Chemical composition was the only physical predictor used during the 
learning process. Suggestions were made for future improvements in terms 
of expanding the database, as well as for other artificial intelligence
calculations. 

\end{abstract}



\begin{keywords}
 \sep heavy fermions \sep deep learning \sep clustering
\end{keywords}

\maketitle

\section{Introduction}

Discovered almost half a century ago, heavy fermions continue to attract
attention of condensed matter physicists 
\cite{coleman07,kirchner20,flouquet05,white15,hfs-book,degiorgi99,stewart84,stewart01,misra-book,onuki-book,coleman15,canfield20}. 
A wide variety of different ground states and
effects have been discovered in them, such as superconductivity, antiferromagnetism, 
ferromagnetism, 
coexistence of superconductivity and antiferromagnetism, quantum critical behavior, 
non-Fermi liquid behavior, etc. Their unusual and intriguing properties 
are believed to originate from hybridization of conduction electrons with localized 
4f or 5f electrons \footnote{It should be pointed out, however, that hybridization 
	is not the only mechanism that can lead to heavy fermion behavior.
    Other routes to effective mass enhancement have also been discussed in the literature 
    such as magnetic frustration, proximity to a quantum critical point, or metal-to-insulator 
    transition \cite{krug03}.}. At high temperatures, hybridization is negligible 
and they behave like 
ordinary metals. However, below the so-called coherence temperature T$^*$,
hybridization causes the development of a correlated state and formation of 
a Fermi liquid with quasiparticles whose effective mass m$^*$ greatly 
exceeds the free electron mass m$_0$, thus giving them their name. 

Heavy fermions are mostly intermetallic compounds which contain
elements with partially filled 4f shells (lanthanides) 
such as Ce, Sm or Yb, or 5f shells 
(actinides) such as U, Np or Pu. However, it should be pointed out that not 
every compound with lanthanide or actinide elements is a heavy fermion. 
Discovering those that are usually relies on the serendipity of a 
researcher.

In this work I propose the use of deep learning to aid the search for 
new heavy fermions. Much of the progress in artificial intelligence 
in recent years has been driven by advances in deep learning using 
artificial neural networks. Deep learning has revolutionized 
certain fields, such as computer vision, self-driving cars, 
natural language processing, drug discovery, 
protein folding, etc \cite{dl-rev-book,lecun15}. This approach has also been 
fruitful in some branches of material science, such as photovoltaics
\cite{gaviria22}, thermoelectrics \cite{iwasaki19},  photocatalysts \cite{singh22}, etc. 
To the best of my knowledge, this so-called {\it in silico} approach to 
materials discovery has not been applied to heavy fermions.
With this work, I make the first attempt at using this novel 
computational technique to aid the search for new heavy fermion 
compounds, as well as to predict their unusual properties.

The paper is organized as follows: In Section \ref{sec-database} I describe the 
database compiled and used for deep learning calculations. Also discussed are
the most important parameters of heavy fermions that are experimentally 
accessible. In Section \ref{sec-clustering} I employ clustering to explore 
the configuration space of exiting lanthanide and actinide compounds, and 
the possibility of finding new heavy fermions. 
Section \ref{sec-deeplearning} starts with a very brief introduction
to deep learning with artificial neural networks, followed
by its application to heavy fermions.  
In Section \ref{sec-summary} the most promising predictions of new 
heavy fermions and their predicted properties are summarized 
and some suggestions for future directions are made.

\section{The database}
\label{sec-database}

Deep learning and other artificial intelligence calculation are often 
referred to as data-driven methods because they (usually) require large amounts 
of input data to train on. During this training process,
the models "learn" various features of the data in a given database, 
which they can then 
use to make predictions about new entries that are not already in the 
database. For example, Google's GoogLeNet was trained
on millions of images and is now able to recognize hundreds of 
different objects or animals \cite{googlenet}.

When attempting deep learning calculations on heavy fermions, one is 
faced with two major problems. First, a comprehensive database
of heavy fermions which are required for deep learning calculations
does not exist. The second 
fundamental problem is that there are relatively few known heavy fermions. 
In order to overcome the first problem, I compiled a database of 
most known heavy fermions, which at the time of writing of this paper
contained more than 200 entries. The second problem will need to be
addressed in the future, as the number of known heavy fermions 
grows. However, I show in this work that even with a limited 
size of input data set,
statistically significant results can be obtained using deep learning. 

Many compounds with lanthanide or actinide elements have been entered in the 
existing materials' databases, such as ICSD, OQMD, COD, etc. However, their
heavy fermion properties have not been recorded. Therefore, for the purpose
of deep learning calculations, I constructed a database of most known 
heavy fermions and their basic physical properties, described 
in more details below. Various literature sources were used, such as 
books, review articles, research articles, web-sites, etc. In addiction 
to chemical composition, every entry in the database included the following
experimentally accessible parameters: coherence temperature T$^*$, 
Sommerfeld coefficient $\gamma$, quasiparticle effective mass m$^*$, 
superconducting critical temperature T$_c$
(for those heavy fermions which are superconducting), Neel temperature T$_N$ 
(for those which are anti-ferromagnetic) and Curie temperature (for those 
which are ferromagnetic). Table~\ref{tab:database} shows the number of entries 
for each  of the lanthanide and actinide elements
in the database. It can be seen that most heavy fermions in the database 
(about 77~$\%$) are either Ce, Yb or U compounds, although there are
also Pr, Nd, Sm, Eu, Np and Pu heavy fermions.

The database will be continuously expanded in the future to include 
newly discovered heavy fermions. It will
also be expanded to include other important parameters that are experimentally
accessible, such as the crystal structure, single ion Kondo temperature (T$_K$), 
hybridization gap ($\Delta$), Debye temperature ($\Theta_D$), 
magnetic susceptibility ($\chi_0$), effective magnetic moment 
($\mu_{eff}$), Kadowaki-Woods ratio, Hall coefficient (R$_H$), etc. 
The database will also be expanded to include not only stoichiometric 
compounds, but also non-stoichiometric heavy fermions (such as CeCu$_{5.5}$Au$_{0.5}$, 
U$_{0.97}$Th$_{0.03}$B$_{13}$, YbCu$_{4.5}$Ag$_{0.5}$, etc). 
Material families closely related to heavy fermions could also be included: 
Kondo insulators (such as SmB$_6$, Ce$_3$Pt$_4$Bi$_3$, YB$_{12}$, etc.), 
d-electron heavy fermions (such as FeSi, CaFe$_4$Sb$_{12}$, Y(Sc)Mn$_2$, etc.), 
oxide heavy fermions (such as LiV$_2$O$_4$, Sr$_3$Ru$_2$O$_7$, Gd$_{0.8}$Sr$_{0.2}$TiO$_3$, etc.), 
heavy fermions in the form of thin films \cite{chatterjee21}, etc. 
In recent years heavy fermions have also been found in quasi-2D systems, such as 
van der Waals heterostructures \cite{vano21}, Moiré bilayers \cite{zhao23} and
van der Waals metals \cite{posey24}. An expanded database
would result in more reliable deep learning calculations and better prediction 
of new heavy fermions and their physical properties.

\begin{table}[t]
\caption{The number of entries in the database at the moment of writing 
of this paper. The number of entries in the COD database \cite{cod} 
is also shown; deep learning is used to search for potential new heavy 
fermions among those COD entries.}
\label{tab:database}
\begin{tabular}{ c|c|c|c }
\toprule
  & Element  & Number of entries & Number of entries in COD  \\ \hline
\multirow{6}{*}{\rotatebox[origin=c]{90}{Lanthanides}}   &  Ce	&     79    & 402  \\ 
      & Pr	&     6     & 255  \\ 
      & Nd	&     6     & 293  \\
      & Sm	&     7     & 263  \\ 
      & Eu	&     6     & 271  \\ 
      & Yb	&    33     & 283  \\ 
  \midrule
\multirow{3}{*}{\rotatebox[origin=c]{90}{Actinides}}       & U	         &   48    & 303  \\ 
      & Np	     &   9     & 49  \\ 
      & Pu	     &   12    & 74  \\  
\bottomrule
\end{tabular}
\end{table}

\subsection{Coherence temperature}

The coherence temperature (also know as the Kondo lattice temperature) 
is one the most important parameters of every heavy
fermion, as it sets the energy scale for the development of coherent
many body state. The symbol used in the literature \footnote{Some authors
\cite{wirth16} distinguish between T$^*$ obtained from magnetic probes and T$_{coh}$ 
derived from  transport measurements. The difference is usually small and will not be discussed here.} 
is either T$^*$ or T$_{coh}$.   
(The coherence temperature should be distinguished from the so-called
single ion Kondo temperature T$_K$, which is a single electron property.) 
The extraction of coherence temperature from experimental data
was discussed in details in Ref.~\cite{yang08}. T$^*$ can be obtained from
a variety of different experimental techniques, such as (magnetic) 
resistivity, optical conductivity, magnetic susceptibility, 
etc. The values of coherence temperature obtained from different techniques
are usually consistent \cite{yang08} within $\pm$~5--10~K. 
	\footnote{Coherence temperatures extracted from 
	ARPES can be 20--50~K higher than the values extracted from the resistivity.}
For most entries in the database the coherence temperature extracted from 
resistivity was used, as those measurements were most readily available. 
It should be noted that at temperature T$^*$ heavy fermions do not undergo 
a phase transition; rather, they {\it crossover} from an incoherent 
state at high temperatures into a low temperatures
hybridized state. Therefore, the changes in all experimentally 
measured quantities are not discontinuous, but rather gradual.

\subsection{Sommerfeld coefficient} 

The Sommerfeld coefficient $\gamma$ quantifies the charge carrier 
contribution in the specific heat
and is proportional to carrier effective mass \cite{kittel-book}. 
In ordinary metals its typical values are between 1--10~mJ/(mol K$^2$), 
which indicates small or negligible enhancement of effective mass compared
to the free electron mass m$_0$. On the other hand, in heavy fermions its values can 
be as high as 100--1,000~mJ/(mol K$^2$), indicating significant mass 
enhancement. Specific heat, along with resistivity, is one of the most basic 
experimental techniques used for studying heavy fermions and values 
of $\gamma$ are readily available for almost all heavy fermions.

\subsection{Effective mass}

At temperatures above T$^*$ the effective mass of charge carriers 
in heavy fermions is small, similar to ordinary metals. However,
as temperature decreases below T$^*$, m$^*$ starts to gradually 
increase, reaching its maximum value at T$\rightarrow$0. Precise 
definition as to what values of effective mass would qualify as 
heavy fermions does not exist 
\cite{coleman07,kirchner20,flouquet05,white15,hfs-book,degiorgi99,stewart84,stewart01,misra-book,onuki-book,coleman15,canfield20}.  
It is assumed that m$^*$ should be at least 10~m$_0$, 
but sometimes even smaller values are referred to as heavy.

Effective mass in heavy fermions (and other materials) 
can be obtained from a variety of different experimental techniques, such as
specific heat measurements (the so-called thermodynamic or thermal effective mass),
ARPES (band effective mass), infrared and microwave spectroscopies
(optical effective mass), de Haas - van Alphen measurements (cyclotron 
effective mass), from upper critical field measurements, thermopower measurements
(Seebeck effective mass), Knight shift measurements, etc. 
In general, the values obtained from different experimental techniques are 
different, in some cases significantly. 
As an example in Table~\ref{tab:upt3} I show 
the values of effective mass for a prototypical and one of the 
most studied heavy fermions, UPt$_3$. As can be seen from the table,
a variety of different values have been reported. Other values, 
not listed here, can be also be found in the literature \cite{joynt02}.

Whenever possible, the values of m$^*$ used in the database
were those from specific heat measurements, as they were most readily available.
Experimentally, the values of thermal effective mass are usually
obtained by comparing the value of $\gamma$ of the Ce-, Yb- or U-based compounds with 
those of the isostructural La-, Lu- or Th-based compounds, respectively. 

\begin{table*}[t]
\caption{Values of effective mass m$^*$ for UPt$_3$ reported in the literature.}
\label{tab:upt3}
\begin{tabular}{ c|c|c }
\toprule
 m$^*$/m$_0$  & Experimental technique & Reference  \\ 
 \midrule
  240	&     Infrared spectroscopy    & Ref.~\cite{sulewski88}  \\ 
 200, 235, 350   &     Microwave spectroscopy   & Ref.~\cite{donovan97}  \\ 
  180   &     Specific heat   & Ref.~\cite{devisser87}  \\ 
 39, 45, 80    &  de Haan - van Alphen effect	&   Ref.~\cite{kimura98}  \\
  25, 40, 50, 60, 90 	&     de Haan - van Alphen effect    & Ref.~\cite{taillefer87}  \\ 
  180   &     Upper critical field   & Ref.~\cite{franse84}  \\ 
  250   &     Point contact spectroscopy   & Ref.~\cite{moser86}  \\ 
\bottomrule
\end{tabular}
\end{table*}

\subsection{Critical temperatures}

Approximately 23~$\%$ of all heavy fermions in the database were superconducting.
Their critical temperatures (T$_c$) were recorded and used as an input 
for deep learning calculations. Based on them, classification predictions were 
made about other possible heavy fermion superconductors. On the other hand,
approximately 44~$\%$ of all heavy fermions in the database were antiferromagnetic.
Their Neel temperatures (T$_N$) were also recorded in the database and based on them, 
classification predictions were made about possible new anti-ferromagnetic heavy fermions. 
It should be added that several heavy fermions were both superconducting and 
antiferromagnetic. Finally, a small number of heavy fermions was ferromagnetic 
and their Curie temperatures were also recorded in the database. However, their number 
was too small for any data driven calculations.

\section{Clustering}
\label{sec-clustering}

The search for new heavy fermion materials was going to be performed among the existing
compounds in the COD database \cite{cod}, which contained about 37,000 
inorganic compounds and alloys. The biggest advantage of this database
was that it only contained materials that had already been synthesized
and their crystallographic properties had been reported in the
literature. For the purpose of deep learning calculations  
only stoichiometric compounds with lanthanide and actinide 
elements listed in Table~\ref{tab:database} were extracted. 
There were 2,161 such entries and their distribution is given in the table.

Before attempting the {\it in silico} search for new heavy fermions, 
it was instructive to explore the very possibility of finding them.
For that task clustering is usually employed. 
Clustering is an unsupervised learning technique used to search for, and 
explore the hidden patterns in the data \cite{cluster-book}. It is 
unsupervised in a sense that it does not require the input data to be
labeled. There are many different clustering algorithms \cite{cluster-book} 
which usually 
produce different results, depending on the input data set and its
structure. Several of them were employed on heavy fermions, and
the so-called DBSCAN \cite{cluster-book} was found to produce the 
most meaningful results.  

In order to visualize the results of clustering, certain dimensionallity 
reduction techniques are usually used along with clustering. 
We previously used the t-SNE to visualize the results of clustering 
of superconductors \cite{roter22}. However, in 
recent years, UMAP \cite{umap} was shown to perform better, 
and was used for heavy fermions.

\begin{figure}[ht]
\vspace*{0.0cm}%
\centerline{\includegraphics[width=4.0in]{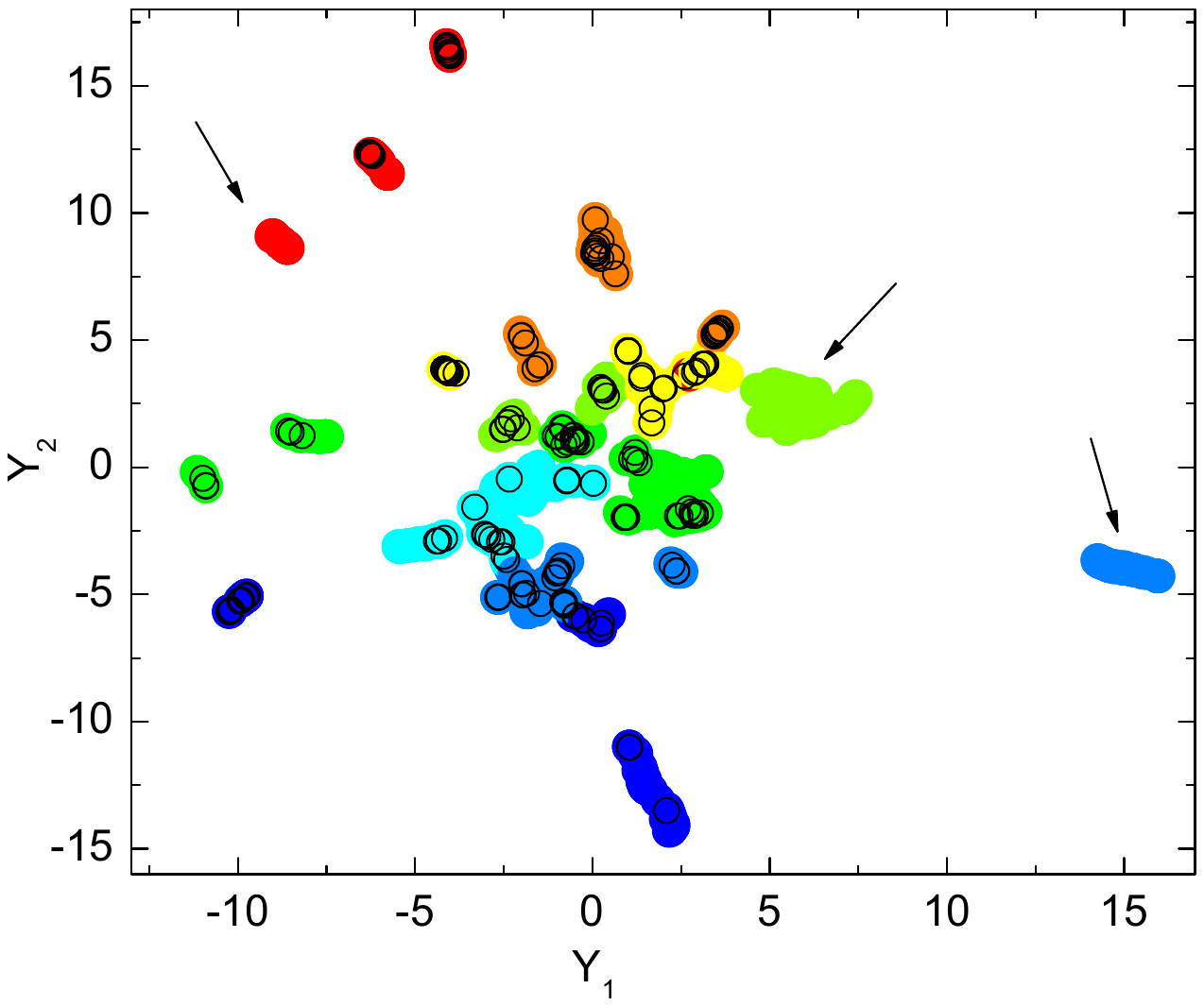}}%
\vspace*{-0.5cm}%
\caption{Clustering of COD compounds and heavy fermions. 
Each compound from COD is represented with a full circle, with 
compounds belonging to different clusters shown with
different colors. Open black circles represent the heavy fermions 
from our database. The arrows point to the clusters which do not
contain any known heavy fermions.}
\vspace*{0.0cm}%
\label{fig:clustering}
\end{figure}

Figure \ref{fig:clustering} shows the results of clustering
using a combination of DBSCAN and UMAP. Note that the two 
axes of the plot do not have any physical meaning \cite{cluster-book}. 
The algorithm was 
able to identify 23 clusters in the COD database, shown
with full circles of different colors. On the other hand,
the heavy fermions from the database are shown with open
black circles. Several conclusions can be drawn from the plot. 
First, heavy fermions from the database do not form any separate 
clusters; instead, they are distributed across different clusters 
of compounds from COD. Second, heavy fermions are conspicuously 
missing from some clusters (marked with arrows). Finally, there
are several clusters which contain only a few heavy
fermions. All these results indicate that there are large areas of 
configuration space
which might contain potential new heavy fermions. In the next 
section I describe how I searched for them using deep learning.


\section{Deep Learning}
\label{sec-deeplearning}

Deep learning is a subset of machine learning that uses artificial
neural networks to extract information ("learn") from the data 
\cite{dl-rev-book,lecun15}. 
Deep learning algorithms have been traditionally employed for two basic supervised 
learning tasks \cite{dl-book}: regression and classification. I employed both of them
on heavy fermions: regression for predicting T$^*$, $\gamma$ and m$^*$,
and classification for predicting their superconducting and 
antiferromagnetic properties.

Deep learning calculations usually require large amounts of data.
Our database of heavy fermions was small, which limited the 
amount of information the networks could learn. However, I showed in 
this work that even with such a small database, statistically significant results 
could be obtained. Because of the small database size, large fluctuations of the 
output were expected. In order to minimize them, every model was trained numerous 
times to create multiple versions and, at inference time, the predictions of the 
different versions of each model were averaged.

\begin{figure}[t]
\vspace*{0.0cm}%
\centerline{\includegraphics[width=3.0in]{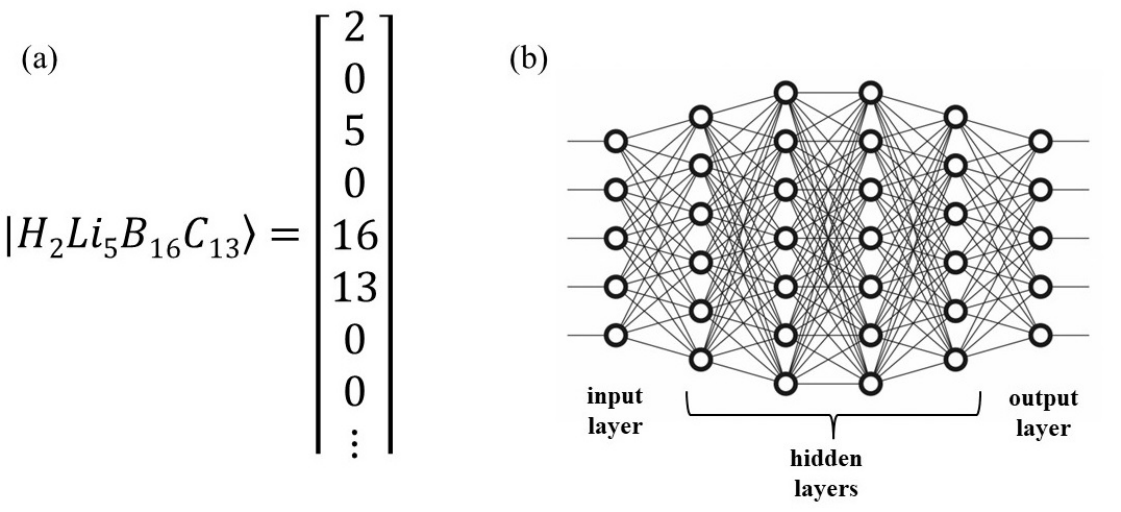}}%
\vspace*{0.0cm}%
\caption{(a) Schematic representation of a hypothetical compound 
H$_2$Li$_5$B$_{16}$C$_{13}$.
The figure shows element-vector representation in the form of 
a column vector (96$\times$1 vector). (b) An example of an 
artificial neural network used in these calculations. 
The input layer was chosen to match the size of the element-vector,
i.e. it had 96 neurons. The number of hidden layers was typically
between 3 and 5, with 50--100 neurons per layer. The output layer
had either one neuron (for regression models) or two neurons
(for classification models with one-hot-encoding).}
\vspace*{0.0cm}%
\label{fig:h2o}
\end{figure}

For deep learning calculations
the chemical formulas of heavy fermions were first represented as 
element-vectors, as was done previously for superconductors \cite{roter20}.
For illustration purposes, in Fig.~\ref{fig:h2o}(a) I show a hypothetical 
compound H$_2$Li$_5$B$_{16}$C$_{13}$
represented as a 96$\times$1 column vector. Note that that the size
of vectors was determined by the heaviest element present in the 
COD database.

Fig.~\ref{fig:h2o}(b) shows schematically a typical neural network
architecture used. The input layer had 96 neurons in order to 
match the size of element-vectors. Different architectures
were tested and it was found that the best performance was achieved
with 2--5 hidden layers, each with 50--100 neurons. Adding more layers
and/or neurons did not lead to improvements, but instead resulted 
in overfitting \cite{dl-book}. Finally, the output layer had either
one neuron for regression calculation, or two neurons for one-hot-encoding
classification. 
The main difference between the networks used for classification
and regression was that for the former ReLU activation function was
used in the output layer, 
whereas for the latter no activation function was used \cite{dl-book}.



For all deep learning calculations the heavy fermion
database was randomly split into training and validation sets, 
usually in the 80~$\%$~--~20~$\%$ proportion.
The network was then trained on the training set for about 100 epochs, 
until satisfactory values of statistical parameters \cite{dl-book} such as 
the loss function, the root-mean-square error (RMSE) or 
the coefficient of determination (R$^2$) were obtained on the validation set.
These trained models were then employed to make predictions about 
possible new heavy fermions among the compounds in the COD database.

\begin{table}[t]
\caption{Predictions of potential new heavy fermions. All compounds listed are from the 
COD database. Also shown in the table 
are their predicted coherence temperature, Sommerfeld coefficient and the effective 
mass m$^*$, as well as predictions whether they are superconducting (SC) 
or antiferromagnetic (AFM).}
\label{tab:predictions}
\begin{tabular}{ l|c|c|c|c|c }
\toprule
 Prediction  & T$^*$ [K]  & $\gamma$ [mJ/(mol K$^2$)]   & m$^*$/m$_0$ & SC & AFM  \\ 
 \midrule
 CeInCu$_3$         & 137    &  270    &   46    &                &  $\checkmark$           \\ 
 YbAuIn$_2$	        & 87 	&  399    &   31    &   $\checkmark$  &  $\checkmark$    \\
 UReRuSi$_2$	        & 44    &  80    &   74    &  $\checkmark$  &                     \\ 
 UZnNi$_4$	           & 150 	&  141    &   56    &                &                \\
 CeCuAgIn 	             & 81 	&  91    &   24     &                &  $\checkmark$    \\
 CePd$_3$As$_2$          & 116   &  261    &   121     &               &                 \\ 
 NdIr$_5$               & 102    &  426    &   154    &    $\checkmark$     &            \\ 
 SmRu$_2$Ge$_2$	        & 167    &  109    &   27    &                &    $\checkmark$     \\ 
 YbPd$_2$Ge$_2$          & 26    &  78    &   44    &                &  $\checkmark$           \\ 
 CeCuSe$_2$ 	             & 122 	&  124    &   78     &                &  $\checkmark$    \\
\bottomrule
\end{tabular}
\end{table}

\subsection{Regression}

\begin{figure}[t]
\vspace*{0.0cm}%
\centerline{\includegraphics[width=4.0in]{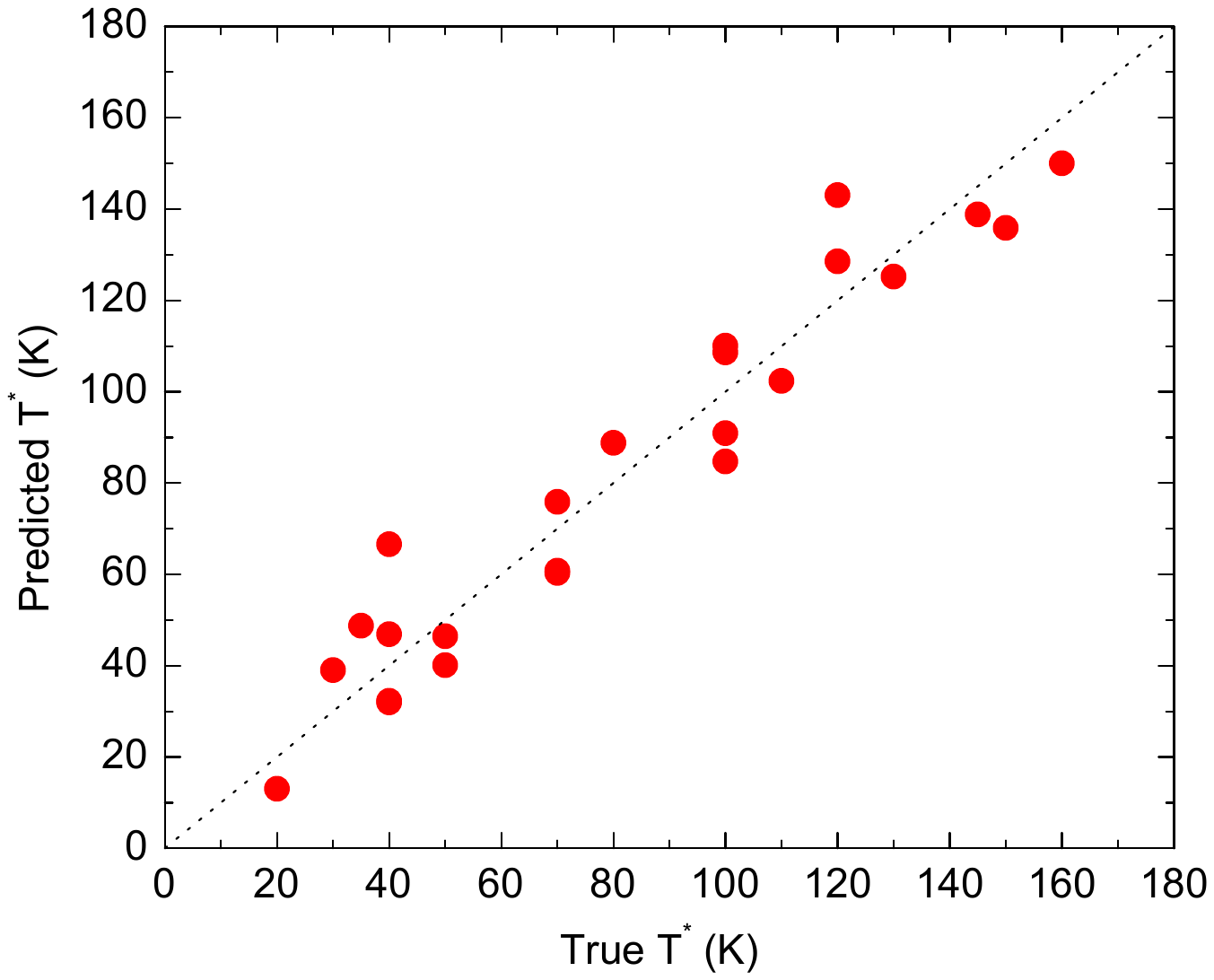}}%
\vspace*{-0.5cm}%
\caption{Regression calculations of coherence temperature T$^*$. 
Similar results were obtained for regression calculations of
Sommerfeld coefficient and effective mass.}
\vspace*{0.0cm}%
\label{fig:regression}
\end{figure}

The goal of regression calculations is to predict the numerical 
values of certain quantities \cite{dl-book}. 
In the case of heavy fermions the goal was to predict
the numerical values of coherence temperature, Sommerfeld coefficient
and effective mass. Training the network with the input data 
for coherence temperature resulted in the RMSE $\simeq$ 10 K, 
and R$^2$ $\simeq$ 78~$\%$. Similar values were 
obtained when training on Sommerfeld coefficient and effective mass. 
These are reasonable values, considering the size of the database. 
Fig.~\ref{fig:regression} shows typical results 
of training for T$^*$.

Once the network was trained and acceptable values of statistical 
parameters were achieved, the network was used to make predictions 
of new heavy fermions, from the COD database. All entries with 
lanthanides and actinides (Table~\ref{tab:database}) were 
represented as element-vectors (Fig.~\ref{fig:h2o}(a)) 
and run through the trained networks,
which calculated their coherence temperature, Sommerfeld coefficient
and effective mass. Those values were then manually inspected and 
the most promising predictions are shown in Table~\ref{tab:predictions}.


\subsection{Classification}

The goal of classification is to predict to which class 
(or group) certain elements belong \cite{dl-book}. 
In the case of heavy fermions the goal was to 
predict whether a given compound was superconducting or not,
and similarly whether a given compound was an antiferromagnetic 
or not. The neural networks used for this task had four 
hidden layers, in addition to the input and output layers. 
The value of R$^2$ obtained for superconductors was 0.88, 
whereas for antiferromagnets R$^2$ $\simeq$ 0.86 was obtained. 
Trained networks were then used to predict whether compounds 
from COD were superconducting or antiferromagnetic. 
The results are shown in Table~\ref{tab:predictions}.

\section{Summary}
\label{sec-summary}

In recent years artificial intelligence techniques have been increasingly
employed to help experimentalists identify potential candidate materials 
before syntheses. In this work I used deep learning calculation 
to aid the search for new heavy fermions. 
In spite of a small input data set, satisfactory values of statistical
parameters were achieved and a number of promising candidates for 
new heavy fermions were identified among the compounds in the COD database.
Their most important physical properties were also predicted, such as 
the coherence temperature, Sommerfeld coefficient and effective mass. 
The potential ground states (superconducting or antiferromagnetic)  
were also discussed. 
Table~\ref{tab:predictions} lists some of the most promising predictions. 
It should be pointed out that none of them were in the training set,
i.e. in the input database. 
Internet search indicates that they have been synthesized in the lab,
but have not been tested for heavy fermion behavior. 
I hope that these deep learning calculations will inspire material 
scientists to measure them. Experimental verification is the ultimate test
for any {\it in silico} search for new materials.



Future calculations would benefit from an expanded database, both 
in terms of the number of entries, as well as other physical parameters. 
An alternative approach in the search for new heavy fermions would be the 
use of generative artificial intelligence, which is currently 
attracting a lot attention \cite{generativeAI,dl-rev-book}. 
We have recently employed it in our search 
for new superconductors \cite{kim22,yuan24} and it has shown 
numerous advantages over conventional methods. Similar approach
could also advance our search for new heavy fermions in the future.

\section{Acknowledgments}

The author thanks Evan Kim and Samuel Yuan for the 
critical comments on the manuscript.

%

\bibliographystyle{cas-model2-names}

\bibliography{hfbib}



\end{document}